\documentstyle{mn}

\input{epsf.sty}

\def\mpc {h^{-1} {\rm{Mpc}}}
\def\and  {\it {et al.} \rm}
\def\rmd {\rm d}

\def\eql#1{\label{eq:#1}}
\def\ec#1{\ref{eq:#1}}
\def\Ec#1{Eq.[\ref{eq:#1}]}
\def\etal{{\rm et~al. }}
\def\we{\hat w}
\def\pe{\hat P}

\def\spose#1{\hbox to 0pt{#1\hss}}
\def\simlt{\mathrel{\spose{\lower 3pt\hbox{$\mathchar"218$}}
     \raise 2.0pt\hbox{$\mathchar"13C$}}}
\def\simgt{\mathrel{\spose{\lower 3pt\hbox{$\mathchar"218$}}
     \raise 2.0pt\hbox{$\mathchar"13E$}}}
\def\be{\begin{equation}}
\def\ee{\end{equation}}
\def\bce{\begin{center}}
\def\ece{\end{center}}
\def\bea{\begin{eqnarray}}
\def\eea{\end{eqnarray}}
\def\ben{\begin{enumerate}}
\def\een{\end{enumerate}}

\def\brr{\begin{array}}
\def\err{\end{array}}

\def\etal{{\rm et~al. }}

\def\nh1{n_{\rm HI}}

\def\p1dk{P_{\rm 1D}(k)}
\def\simlt{\mathrel{\spose{\lower 3pt\hbox{$\mathchar"218$}}
     \raise 2.0pt\hbox{$\mathchar"13C$}}}
\def\simgt{\mathrel{\spose{\lower 3pt\hbox{$\mathchar"218$}}
     \raise 2.0pt\hbox{$\mathchar"13E$}}}

\begin{document}

\title{Inverting the Angular Correlation Function}

\author[S. Dodelson and E. Gazta\~{n}aga ]{
Scott Dodelson$^{1,2}$ and  Enrique Gazta\~{n}aga$^{3}$
\vspace{1mm}\\
$^1$ NASA/Fermilab Astrophysics Center, P.O. Box 500, Batavia, IL 60510 USA \\
$^2$ Department of Astronomy and Astrophysics, University of Chicago, Chicago, IL
60637 USA\\
$^3$ Consejo Superior de Investigaciones Cient\'{\i}ficas (CSIC), 
Institut d'Estudis Espacials de Catalunya (IEEC), \\
Edf. Nexus-201 - c/ Gran Capitan 2-4, 08034 Barcelona, SPAIN}

\maketitle 
 
\def\mpc {h^{-1} {\rm Mpc}}
\def\impc {h {\rm Mpc}^{-1}}
\def\and  {{\it {et al.} }}
\def\rmd {{\rm d}}

\begin{abstract}

The two point angular correlation function is an excellent
measure of structure in the universe. To extract from it the
three dimensional power spectrum, one must invert Limber's
Equation. Here we perform this inversion using a Bayesian prior
constraining the smoothness of the power spectrum.
Among other virtues, this technique allows
for the possibility that the estimates of
the angular correlation function are correlated from bin to bin.
The output of this technique are estimators for the
binned power spectrum and a full covariance matrix. 
Angular correlations mix small and large scales but
after the inversion, small scale data can be trivially eliminated, 
thereby allowing for
realistic constraints on theories of large scale structure. We analyze
the APM catalogue as an example, comparing our results with previous
results. As a byproduct of these tests, we find -- in rough agreement with
previous work -- that APM places stringent constraints
on Cold Dark Matter inspired models, with the shape parameter constrained to
be $0.25\pm 0.04$ (using data with wavenumber  $k \le 0.1 h{\rm Mpc}^{-1}$).
This range of allowed values use the full power spectrum covariance matrix, 
but assumes negligible covariance in the off-diagonal angular correlation 
error matrix,  which is estimated  with a large angular resolution 
of $0.5$degrees (in the range  $0.5$ and $20$ degrees). 

\end{abstract}


\section{Introduction}

The two point angular correlation function, $w(\theta)$, has
emerged as one of the most powerful measurements in cosmology. It
is constructed from photometric catalogues, bypassing 
the need to take time-consuming
redshifts. Photometric catalogues contain many more galaxies
 than do redshift surveys. This advantage in statistics is often 
enough to offset the extra information about distance conatined
in redshift surveys. In fact, one might argue that the redshift is
prone to misinterpretation due to redshift-space distortions caused
by peculiar velocities. Angular catalogues which implicitly integrate
over all line-of-sight distances avoid these distortions. These
general arguments are backed up by the current state of affairs
regarding the three dimensional power spectrum. The surveys which
constrain the power spectrum on the largest scales are the APM (Maddox
\etal 1990) and EDSGC (Collins, Nichol, and Lumsden 1992) surveys, 
each of which contain angular 
positions for well over a million galaxies.

Extracting the power spectrum from $w(\theta)$ 
requires an inversion of Limber's
Equation which describes how the angular correlation function probes
 structure on all scales. This a classical example of
an inverse (linear) problem. Schematically, 
Limber's Equation reads
\be
w_i = K_{i\alpha} P_\alpha
\eql{LimberSch}
\ee
where Roman indices label angular bins, while Greek indices label
bins in $k-space$. Here,
we use the same kernel $K$ used by Baugh \& Efstathiou (1993,1994)
and Gazta\~{n}aga \& Baugh (1998). These groups
have successfully extracted the power spectrum from the APM
$w$ by using Lucy's method to invert \Ec{LimberSch}. Roughly, this
entails fitting a function to the observed $w$, and then iterating
until one finds the best power spectrum. We say ``successfully''
because Gazta\~{n}aga \& Baugh (1998)
 have tested their method on simulations. They show that
the power spectrum they obtain by measuring $w(\theta)$ in the simulations
and then inverting agrees well with the true power spectrum. 

Over the coming decade, a wide variety of galaxy surveys and cosmic microwave
background anisotropy experiments will generate a wealth of data for cosmologists
to study. It is conceivable that we will enter the age of precision cosmology, 
where we strive to determine fundamental physical and cosmological parameters
extremely accurately. To be effective in this environment a given measurement
must provide not only a reliable estimator for the quantity under consideration
(e.g. the power spectrum) but also a reliable error matrix. The purpose of this
paper is to introduce a technique which extracts from $w(\theta)$ not
only an estimator for the power spectrum $P(k)$, but also a reliable
measure of the error matrix, $C_P$. This error matrix provides a quantitative
measure of the accuracy of the estimator. The fact that it is a matrix,
and not just a vector of diagonal error bars, allows for the very real
possibility that the power spectrum estimator in nearby bins will be correlated.
This set of $(P,C_P)$ can then be used to constrain cosmological models
and parameters.

The question naturally arises as to why one extracts the power spectrum from
the angular correlation function; why not simply constrain
models with the angular data? Section 2  provides an answer to this question.
Section 3 describes the technique, which is very similar to that
discussed in Press \etal (1992). Since we want to test not only
our estimates of $P(k)$ but also our estimates of the error matrix, 
it is not sufficient to compare extracted $P$ with the true $P$. Section
4 describes the way we will test the error matrix. Finally, section
5 presents the results for the APM survey. In the process, we will
obtain a set of powerful constraints on Cold Dark Matter (CDM) models.
We conclude in section 6 with suggestions for future work.

\section{Why Invert?}

There are several reasons why an extracted three dimensional power
spectrum is more useful than the two point angular correlation
function. The kernel in \ec{LimberSch} depends on details
of the survey: how deep does it go? What is its selection function?
Therefore, $w(\theta)$ is very dependent on the survey. Different surveys
can and do get different results. The power spectrum, on the other hand,
is more closely tied to theories; in principle the estimators
constructed for the power spectrum try to be as unbiased as possible.

The most important consideration, though, has nothing to do with these
issues of taste. Rather, the most important issue is that the angular
correlation function depends on physics about which we know little:
the physics of nonlinearities and hydrodynamics which operate on small
scales. Figure 1 illustrates this point. It shows $w(\theta)$ for several
different angles -- all thought to be ``large'' -- as a function of the maximum
value of $k$ considered in the sum in \Ec{LimberSch}. That is, we perform
the integral in \Ec{LimberSch} 
 up to $k=k_{\rm max}$ to find $w(\theta,k_{\rm max})$. 
We see that even these large angles get non-negligible contributions
from relatively large wave numbers (small scales). For example, even $w(5^\circ)$
depends somewhat on the power between $k = 0.1 \rightarrow 0.2 h {\rm Mpc}^{-1} $.
These are scales small enough to be affected by nonlinearities and hydrodynamics.
The other side of the coin is that $w(1^\circ)$, while dominated by small scale
power, does contain some information about large scales, and it would be a shame to 
throw this information away. 

\begin{figure}
\vspace{-2.5 cm}
\centerline{\epsfxsize=8.truecm \epsfbox{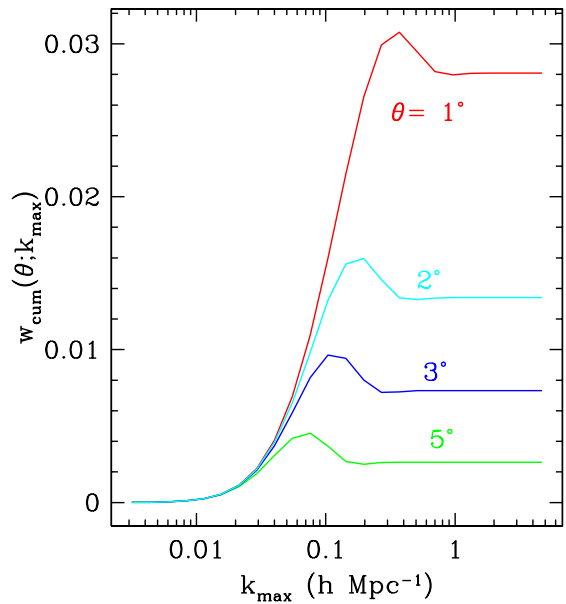}}
\caption[Fig1]{\label{wofk}  The cumulative value of the angular correlation
function, considering all $k-$ modes up to $k_{\rm max}$. The response function is
that for the APM survey, and the power spectrum used is the best-fit 
CDM-like power spectrum for
that survey.}
\end{figure}

Inversion offers an excellent way out of this dilemma. Upon inverting, we will
be using all the large scale information and throwing out all the small
scale information. Mathematically, this is done in a trivial way: 

\begin{itemize}

\item Perform the inversion and get $P$ and its associated error matrix $C_P$.

\item Throw away all the small scale bins from step 1 by truncating $P$
and $C_P$. This is equivalent to marginalizing over these modes. That is, the
resulting smaller error matrix has implicitly integrated over
all possible values of $P$ on
small scales. It is not contaminated at all by small scale information.

\item Use these large scale modes to constrain theories.

\end{itemize}

We now proceed to outline the inversion process.

\section{Inversion} 

One's first thought upon encountering \Ec{LimberSch} is to simply invert
the matrix $K$ to obtain a estimator for the power spectrum:
\be
\pe_\alpha = K^{-1}_{\alpha,i} \we_i
\eql{SimpleMinded}\ee
where the set of ${\we_i}$ are the estimators for the angular correlation
function. This simple approach does not work. The inversion
is very unstable and typically leads to nonsense. This is because
the solution is degenerate (or $K$ is singular) as we typically 
want to obtain more information about $P(k)$ than available in $w(theta)$.
In order to regulate the
inversion, we need to introduce a bit of formalism.

We will assume that we are handed a set of $N_w$ estimators for the angular
correlation function, ${\we_i}$, each $i$ corresponding to one of the $N_w$ angular
bins. In addition, we assume we are handed the full error matrix for 
this set of estimators, $C_w$, an $N_w \times N_w$ matrix. 
This could be computed from first principles, or it could be estimated
from a set of simulations. Mathematically, the simplest assumption
is that the errors have a Gaussian distribution so that the probability
that the angular correlation function is equal to ${w_i}$ is
\be
{\cal P}\left( \bigg\{w_i\bigg\} \right) 
\propto
\exp\Big\{
-{1\over 2} \left( w_i - \we_i \right)
(C_w^{-1})_{ij} \left( w_j - \we_j \right)
\Big\}
\eql{dist}\ee

In order to invert, we need to assume
more, we need to assume that the power spectrum
is ``smooth.'' This assumption is put in by a second
exponential in the probability distribution (Press \etal, 1992):
\bea
{\cal P}\left( \bigg\{P_\alpha\bigg\} \right)
&\propto&
\exp\Big\{
-{1\over 2} \left( K_{i\alpha}P_\alpha  - \we_i \right)
C^{-1}_{ij} \left( K_{j\beta}P_\beta - \we_j \right)
\Big\}
\cr
&\times&\exp\Big\{
-{\lambda\over 2}
P_\alpha H_{\alpha\beta} P_\beta
\Big\}
\eql{distp}\eea
where now we have explicitly eliminated $w$ in favor of $P$
using Limber's Equation.
The second exponential here can be viewed
 as a {\it prior} distribution. It is implemented
by putting in some matrix $H$ (see Press \etal 1992 for some
examples) which makes it costly for $P$ to vary too much. 
Here we use for $H$  the first difference matrix, equation 18.5.3. 
in Press et al. (1986). We have tried other difference
matrices with little effect on the results.
For historical reasons, we have actually taken $k P(k)$ to be the unknown
function. This means that we are smoothing $k P(k)$ to be locally flat.

This entire $H$ term is weighted by a free parameter $\lambda$,
which allows one to tune the relative weights of the
first and second exponential. One of the questions which will
occupy us below is, What should $\lambda$ be set to?
If one distrusts priors,
$\lambda$ should be set very small to have little effect.
This will have to be balanced by the requirement that the
inversion is stable.

The argument of the exponential in \Ec{distp}
is quadratic in $P$. Thus it can be rewritten as
\be
{\cal P}\left( \bigg\{P_\alpha\bigg\} \right)
\propto
\exp\Big\{
-{1\over 2} \left( P_\alpha  - \pe_\alpha \right)
(C_P^{-1})_{\alpha\beta} \left( P_\beta - \pe_\beta \right)
\Big\}
\eql{distfin}\ee
where the estimator for the power spectrum is
\be
\pe_\alpha \equiv  (C_P^{-1})_{\alpha\beta}
(K^t)_{\beta i} (C_w^{-1})_{ij} \we_j
\ee
and the error matrix is
\be
C_{P,\alpha\beta} \equiv 
\left[
(K^t)_{\alpha i} (C_w^{-1})_{ij} K_{j\beta} + \lambda H_{\alpha\beta}
\right]^{-1}.
\ee
Note that in the limit that $\lambda \rightarrow 0$ we do indeed recapture our
initial guess, $\pe = K^{-1} \we$. By varying $\lambda$ we can now
move away from this unstable solution. Also note that while Press \etal
(1992) refrain themselves from talking about the $\lambda P H P$
term as a prior, this interpretation is essential if we are to 
obtain an error matrix for $P$.

\section{Tests of Inversion: Constraints on Cosmological Models}

The simplest test of an inversion technique is to compare the recovered
power spectrum with the true, underlying spectrum. There are
several problems with this, though. First, and foremost,
we do not know the true power spectrum. This difficulty can
be surmounted by working with simulations, with which it
is possible to generate angular catalogues. However, even
if the true power spectrum is known, there is always the
quantitative problem of determining how good the inversion
is. This problem is exascerbated when the estimates of
the power spectrum in adjacent bins are correlated. 
How do we make sense of the full error matrix for $P(k)$?

One way to test inversion is to put constraints on
the parameters in a cosmological model. First, the
constraints can be placed in the parameter space using the
angular data, and then a second set of constraints can be
drawn using the inferred power spectrum. These should agree.
In fact, the method described in \S 2 is linear: the
estimators for $P(k)$ are linear in the estimators of
$w$. Really, then, the inversion process can be thought of
as a new basis for the data, the ``$P$'' basis instead of
the ``$w$'' basis. Through this prism, it is clear that 
the constraints on parameters should be independent of
basis. 

\begin{figure}
\centering
\vspace{-2.5 cm}
\leavevmode\epsfxsize=8cm \epsfbox{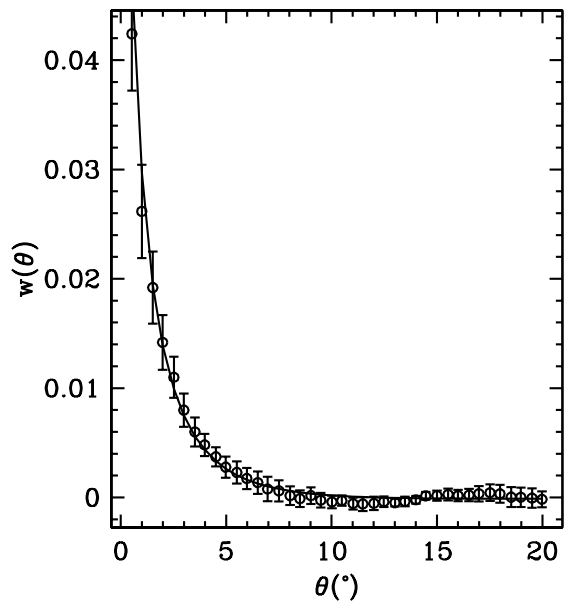}  
\caption[Fig2]{\label{w}  The angular correlation function from the APM
survey (points with error bars). Also shown is the angular correlation function
associated with the best fit CDM model from the contours in Figure 3.}
\end{figure}

For our example, we will choose the APM survey
(Maddox \etal 1990), with data points and error bars 
shown in Figure 2. Errors are from the dispersion in
4 subsamples of the APM pixel maps
(same as in Baugh \& Efstathiou 1994)
and are assumed to be diagonal. 
We have tried several ways of estimating the full 
covariance matrix for $w$. First, we have estimated it directly
from the four quadrants of the APM survey. 
This produced a covariance matrix which was far
too noisy. We then generated ten separate mock APM
catalogues and estimated the covariance matrix from these
forty sets ($10\times$ four quadrants for each). This
proceedure worked on smaller angular scales but broke
down on the largest scales (probably because the simulations
were not large enough). One could also estimate $C_w$ analytically
by assuming that the underlaying fluctuations are Gaussian.
The results that we have obtained using the full covariance
matrix of $C_w$, but with smaller angular scales (ie $\theta < 10$ deg),
give larger errors, but not much different in qualitative terms
from the results that will be present here.
This paper is mostly concerned
with the inversion process for fixed ($w,C_w$) so 
we will simply assume here that $C_w$ is diagonal and leave 
off-diaginal errors for future work.

The model we choose to constrain is the CDM-like power spectrum with
power spectrum
\be
P(k) = A k T^2(k;\Gamma)
\ee
where $T$ is the BBKS (Bardeen, \etal 1986) transfer function. 
There are two free parameters in this model: the amplitude $A$
and the shape parameter $\Gamma$. We first determine the constraints
on these parameters using the angular correlation data. Specifically, we
calculate
\be
\chi^2 \equiv \sum_{i=1}^{N_w}  {(w_i(\Gamma,A) - \we_i)^2
			\over C_{w,ii} }
\ee
where we have explicitly indicated that $w_i$ depends on the 
parameters $(A,\Gamma)$ via Limber's Equation \Ec{LimberSch}. Figure
3 shows the allowed one- and two- sigma regions in this parameter space.
These contours will be our basis for judging the efficiency of the
inversion. If we include all $k-$ modes, we should recover
identical contours from the inverted spectrum. The main advantage of
inversion is that once we have performed a successful inversion, we
can throw out the small scale $k-$ modes and generate a new, more
reliable allowed region.

\begin{figure}
\centering
\leavevmode\epsfxsize=7cm \epsfbox{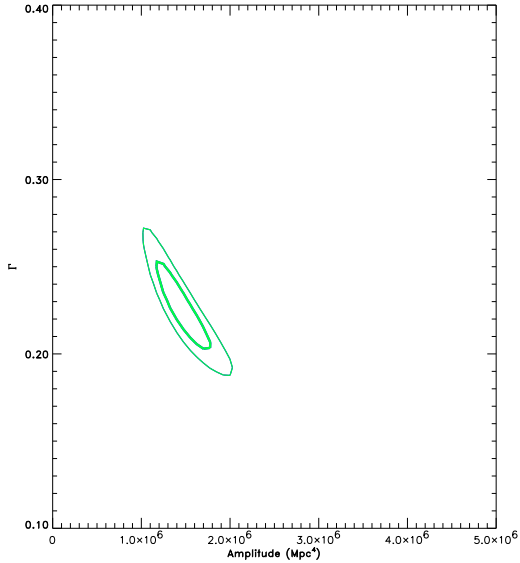}
\caption[Fig3]{\label{chicont}  Allowed one- and two- sigma regions in
CDM-like parameter space from APM survey. ``Standard CDM'' corresponds to 
$\Gamma = 0.5$; COBE-normalization sets $A \simeq 10^7$ Mpc$^{4}$.}
\end{figure}

\section{Inversion of APM Correlation Function}

We now test the inversion algorithm of \S 3 on the APM data.
The extracted power is shown in Figure 4. The error bars are
the square roots of the diagonal elements of $C_P$. For each bin, this 
error then includes the uncertainties induced by marginalizing over
all other $k-$ modes. Also shown in Figure 4 is the power spectrum 
obtained by Lucy's method 
(from Table 2 in Gazta\~{n}aga \& Baugh 1998).
These agree very well except on small
scales. However, most of the disagreement is illusory
because we are only using angular scales $\theta >0.5$ degree which
has limited information about the power on scales $k > 0.5 {\rm Mpc}^{-1}$.

\begin{figure}
\centering
\vspace{-2.5 cm}
\leavevmode\epsfxsize=8cm \epsfbox{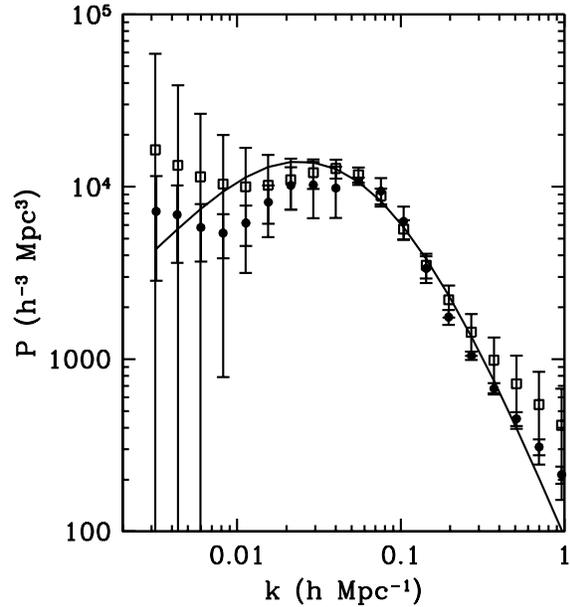} 
\caption[Fig4]{\label{lambone} The power spectrum obtained by inverting
the APM correlation function. Square symbols are from Bayesian inversion
described in \S 3; circles are from Lucy's method. The free parameter $\lambda$
in \Ec{distp} has been set to an ``intermediate'' value here $10^{-4}$. The solid curve
corresponds to the best-fit CDM like model.}
\end{figure}

Some of the disagreement between the two methods on small scales 
results from a more subtle effect. The estimates of the
power spectrum on small scales are highly correlated, as shown in Figure 5. 
This means that
the overall amplitude in any of these modes is uncertain\footnote{One way
to see this is to add to a $2\times 2$ identity matrix a very large number  
to each of the matrix elements (including the off-diagonal ones). Upon diagonalizing
this matrix, one sees that one eigenmode -- the sum of the two original ones -- has
a huge eigenvalue, while the eigenmode corresponding to the difference has eigenvalue
equal to one. In fact, analysts of cosmic microwave background experiments
often add a very large number to every element of the covariance
matrix to account for the fact that the average is unknown (Bond, Jaffe, \& Knox, 1998).}
This uncertainty
is reflected in the large diagonal error bars. One can think of this in terms of
linear combinations of the modes. One linear combination -- the sum of power in
the all the bins or the average power -- is very uncertain. However, all other
linear combinations -- and therefore the shape -- are known quite accurately.

\begin{figure}
\centering
\leavevmode\epsfxsize=8cm \epsfbox{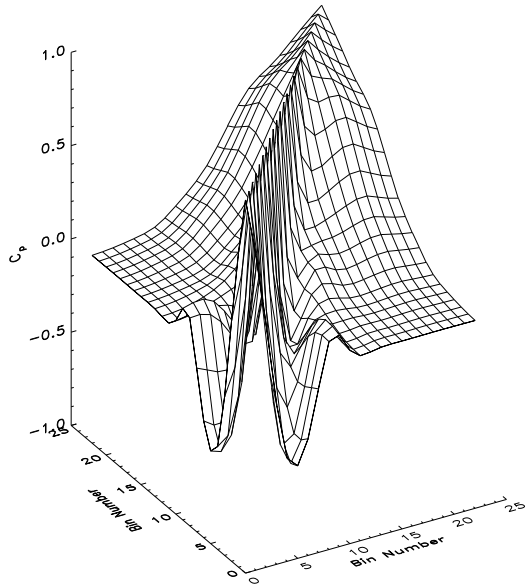}
\caption[Fig5]{\label{pcovone} The full covariance matrix of the power spectrum
depicted in Figure 4. Each element is normalized by the diagonal: $C_{P,ij}/
\sqrt{C_{P,ii} C_{P,jj} }$. Note the large covariances among bins $15-23$,
corresponding to $k \ga 0.2 h {\rm Mpc}^{-1}$.}
\end{figure}

We now test the inversion to see if it recaptures the constraints on 
the parameters $\Gamma$ and $A$ that were obtained in Figure 3 the 
angular correlation function itself. Figure 6 shows these two sets of
contours; they agree extremely well, suggesting that the inversion has
been successul.

\begin{figure}
\centering
\leavevmode\epsfxsize=7cm \epsfbox{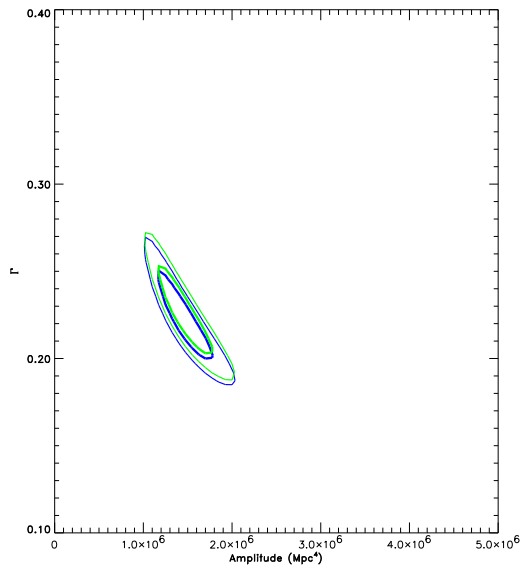}
\caption[Fig6]{\label{chipone} Allowed one- and two-sigma region in parameter
space from APM using the angular correlation function directly and the extracted
power spectrum with its error matrix $C_P$.}
\end{figure}

Finally, Figure 7 shows the contours one obtains by throwing out
the small scale data ($k <= 0.1 h {\rm Mpc}^{-1}$). As one might
expect, the allowed region gets much larger, but the qualitative statement
that small $\Gamma\sim 0.25$ is preferred remains true.
Gaztanaga \& Baugh (1998) found a higher value of $\Gamma = 0.45 \pm 0.10$,
over the four points in range $0.02<k<0.06  h {\rm Mpc}^{-1}$. For this
narrower range of scales our allowed region 
does indeed peak closer to $\Gamma =0.4$ as expected from the agreement in
both $P(k)$ estimations shown in Figure 4.

\begin{figure}
\centering
\leavevmode\epsfxsize=7cm \epsfbox{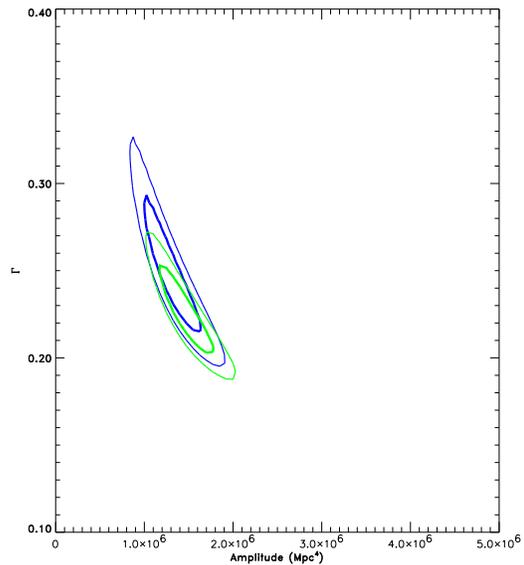}
\caption[Fig7]{\label{chiplowk} Same as Figure 6, except now only the
large scale data has been used.}
\end{figure}

Figures 4-7 are for a particular value ($10^{-4}$) of the free parameter $\lambda$
which sets the importance of the smoothness prior in \Ec{distp}. What motivates
this choice and how do the results change as $\lambda$ gets bigger or smaller?

First consider reducing $\lambda$ thereby trying to eliminate the dependence
on the smoothing prior. Figure 8 shows the inferred power spectrum in this
case. Although the mean $P(k)$
 agrees with that using a stronger smoothness prior, the error
bars are significantly larger. 
The larger errors result from the fact that the power
spectrum estimates are much more highly correlated if the prior is weak, 
due to the degeneracy in the inverse problem solution.
To understand this, consider the
limit of no smoothness prior. In that case, it is possible to fit the $w(\theta)$
data with an extremely choppy power spectrum. 
One bin might have huge power, while another
has negative power. This choppiness is evident on large scales in Figure 8. 
One could still try to use such a choppy power spectrum to
fit models. But the results in each bin do not make much sense by themselves
and there are
large covariances among the different bins. So this is not a very
useful representation of the data.
The large correlations between different bins also affects
the constraints on CDM models, which use 
only the large scale data. This is  shown in Figure 9. 
The small $\lambda$
constraints (weak prior) are much less restrictive than the stronger prior. This reflects
the fact that the large scale estimates are highly dependent on the small scale estimates.
Hence, marginalizing over the latter leads to constraints which are not
very tight. Incidentally, the constraints obtained using all the data are identical
to the constraints coming from $w(\theta)$ itself (the analogue of Figure 6). So
the inversion is accurate, but not useful because the modes are so correlated.

\begin{figure}
\centering
\vspace{-2.5 cm}
\leavevmode\epsfxsize=8cm \epsfbox{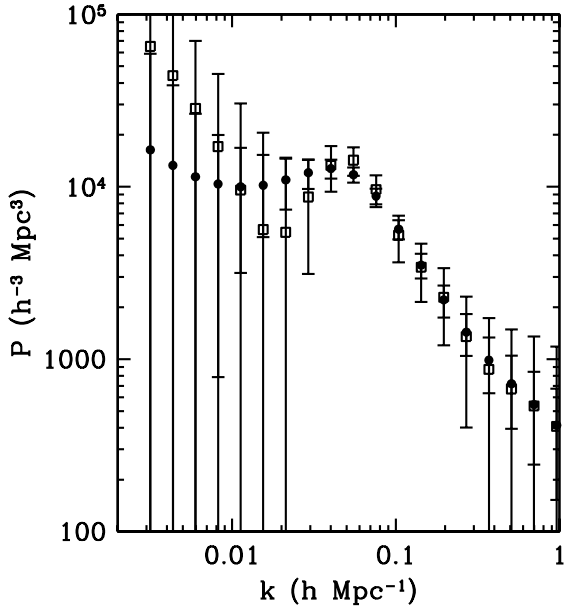} 
\caption[Fig8]{\label{p4v5} The inverted power spectrum using a weak smoothness
prior ($\lambda = 10^{-5}$; squares) and the moderate prior ($\lambda = 10^{-4}$; circles)
used previously. The larger error bars in the weak prior case result from higher
bin-to-bin correlations.}
\end{figure}

\begin{figure}
\centering
\leavevmode\epsfxsize=7cm \epsfbox{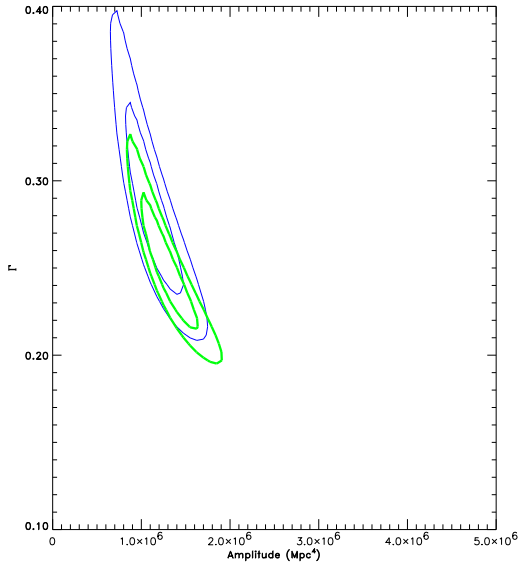}
\caption[Fig9]{\label{chi4v5} Constraints on parameter space using only large
scale data from moderate prior (thick lines) and weak prior (thin lines). The
weak prior leads to less restrictive constraints.}
\end{figure}

Introducing a stronger prior leads to an inaccurate power spectrum extraction. 
The prior is given too much weight, and the data loses out. We can see this
in Figure 10, which shows the power spectrum using a strong prior as opposed
to the moderate one advocated earlier. At intermediate scales the power estimates
differ decidedly. Examining the constraints on the cosmological parameters
illustrates that the incorrect inversion is the one using the strong prior.
Figure 11 shows that the strong prior leads to incorrect constraints. 
The information in the data has not been processed accurately. 

\begin{figure}
\centering
\vspace{-2.5 cm}
\leavevmode\epsfxsize=8cm \epsfbox{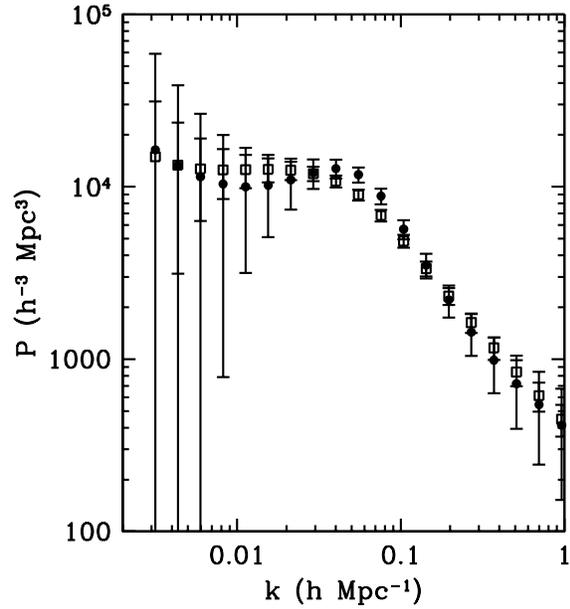} 
\caption[Fig10]{\label{p4v3} The inverted power spectrum using a strong smoothness
prior ($\lambda = 10^{-3}$; squares) and the moderate prior ($\lambda = 10^{-4}$; circles)
used previously. The strong prior leads to an incorrect determination of the power
on intermediate scales.}
\end{figure}

\begin{figure}
\centering
\leavevmode\epsfxsize=7cm \epsfbox{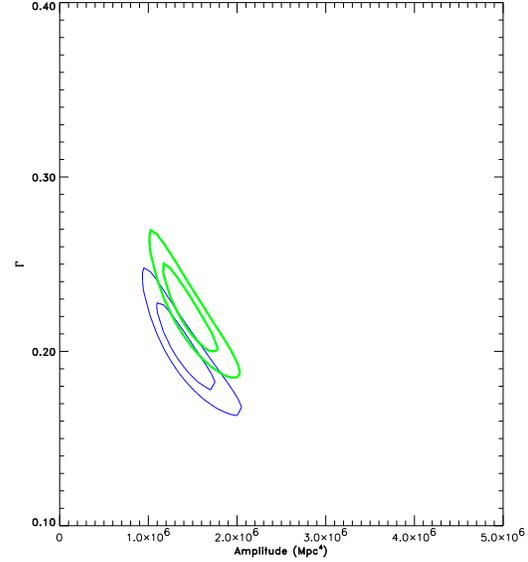}
\caption[Fig11]{\label{chi4v3} Constraints on parameter space using all
the data from moderate prior (thick lines) and strong prior (thin lines). The
strong prior leads to incorrect constraints, while the moderate prior reconstructs
perfectly the constraints obtained from $w(\theta)$ directly.}
\end{figure}

\section{Conclusions}

Very valuable information is contained in the angular correlation function.
A useful way to extract this information is to invert it and obtain an
estimator for the three dimensional power spectrum. We have introduced here a 
method that is different from the one used previously and have focused intensely
on its advantages and its features. However, it is important not to lose sight
of the fact that this inversion technique agrees extremely well with Lucy's method,
the previous inversion tool. This agreement suggests that we (as a community) are
correctly inferring the power spectrum from the angular correlation function.

We worked with the angular APM data; our results for the three dimensional
power spectrum and its error matrix are shown in Figures 4 and 5. Files
containing these numbers are available at
http://www-astro-theory.fnal.gov/Personal/dodelson/Inversion/power.html.

Having reiterated these successes, we warn that
there are a number of issues not explored here that warrant further study:

\begin{itemize}

\item The distribution for the angular correlation function is
not Gaussian, as assumed in \Ec{dist}. Indeed, even if the fluctuations
are Gaussian -- as they are predicted to be in inflationary models on large
scales but are certainyl not on small scales -- the likelihood
function is not Gaussian in $w$. The full likelihood function is
hopelessly complicated (Dodelson, Hui \& Jaffe, 1997), 
but perhaps there are approximations that can
be made which account for the non-Gaussianity. Indeed, there has recently
been some progress along these lines (Meiksin \& White, 1999; Scoccimarro,
Zaldarriaga \& Hui, 1999; Hamilton, 1999; Hamilton \& Tegmark, 1999).

\item In performing the inversion, we assumed that the power spectrum
was separable: $P(k,t) = P(k) f(t)$ and assumed a simple form for $f(t)$ (
$f = a^\beta$).  Similarly, we
have not explored at all the uncertainties in the selection function. We have also
used the small angle approximation. 
All of these fold into the kernel in Limber's Equation
\Ec{LimberSch}. They may be sufficient for APM (e.g. see 
Gazta\~{n}aga \& Baugh, 1998),  but need to be revisited
for deeper surveys, such as the Sloan Digital Sky Survey.

\item We assumed that the covariance matrix for the angular correlation
function is diagonal. The exact nature of this matrix depends on the binning
proceedure, but clearly it is not diagonal. 
Our efforts to obtain the full covariance
matrix from simulations failed, but perhaps simulations on small scales could be
supplemented by linear calculations on large scales to obtain the full covariance
matrix.

\item Related to the first and third points is our assumption that 
the covariance matrix $C_w$
does not depend on $w$ itself. This again is not true and needs to be accounted for
when constraining parameters in a cosmological model.

\item Although we explored the consequences of varying the smoothness
prior, we did not explore how these variations couple to: (i) different
binning schemes for $w(\theta)$; (ii) different binning schemes for $P(k)$;
or (iii) theoretical models which vary more rapidly than the ones
discussed here (e.g. high baryon models retain signatures of primordial
acoustic oscillations).

\end{itemize}

All of these assumptions were implicit in previous inversions, 
and other ways of obtaining
the power spectrum involve a similar number of assumptions. So measuring
the angular correlation function
still is an excellent way to get at the power spectrum. Clearly, though,
more work is needed to enable the extraction to be as 
powerful and accurate as possible.

{\bf Acknowledgments}
We thank Carlton Baugh for helpful comments.
We thank the referee for his inciteful comments.
His careful reading of the manuscript has really
helped us improve the text. We are especially grateful to him
for taking the time to work through our arguments, and
for spotting some of the weaknesses in them.
We acknowledge support from NATO
Collaborative Research Grants Programme CRG970144.
SD is supported by NASA Grant NAG 5-7092 and the DOE.
EG acknowledges support by spanish DGES(MEC), 
project PB96-0925.

\section{References}
\def\refe {\par \hangindent=.7cm \hangafter=1 \noindent}
\def\apj { ApJ }
\def\astroph{{\tt astro-ph/}} 
\def\aap {A \& A }
\def\ajs{ ApJS }
\def\aj{AJ}
\def\prd{Phys ReV D}
\def\apjs{ ApJS }
\def\mnras { MNRAS }
\def\apjl { Ap. J. Let. }
\refe Baugh, C.M., Efstathiou, G., 1993, \mnras 265, 145 
\refe Baugh, C.M., Efstathiou, G., 1994, \mnras 267, 323
\refe Bardeen, J.~M., Bond, J.~R., Kaiser, N., \& Szalay, A.~S.,
1986, \apj, 304, 15
\refe Bond, J.R., Jaffe, A.~H., \& Knox, L., 1998, \prd, 57, 2117
\refe   Collins, C.~A.
Nichol, R.~C., \& Lumsden, S.~L. 1992, \mnras, 254, 295
\refe Dodelson, S., Hui, L., Jaffe, A.~H., 1997, \astroph 9712074
\refe Gazta\~{n}aga, E. \& Baugh, C.M., 1998, \mnras, 294, 229 
\refe Hamilton, A.J.S., 1999, \astroph 9905191
\refe Hamilton, A.J.S. \& Tegmark, M., 1999, \astroph 9905192
\refe  Maddox, S.~J., Efstathiou, G.,
Sutherland, W.~J., \& Loveday, L. 1990, \mnras, 242, 43P
\refe Meiksin, A. \& White, M., 1998, \astroph 9812129
\refe Press, W.H., Teukolsky, S.A., Vetterling, W.T., \& Flannery, B.P.,
	1992,	{\it Numerical Recipes} (Cambridge)
\refe Scoccimarro, R., Zaldarriaga, M. \& Hui, L., 1999, \astroph 9901099

\end{document}